\documentclass[pt12, a4paper]{article} 

\usepackage{graphicx}
\usepackage{url}
\usepackage{placeins}
\usepackage[hidelinks]{hyperref}

\usepackage{tikz,xcolor,hyperref}
\definecolor{lime}{HTML}{A6CE39}
\DeclareRobustCommand{\orcidicon}{
	\begin{tikzpicture}
	\draw[lime, fill=lime] (0,0) 
	circle [radius=0.16] 
	node[white] {{\fontfamily{qag}\selectfont \tiny ID}};
	\draw[white, fill=white] (-0.0625,0.095) 
	circle [radius=0.007];
	\end{tikzpicture}
	\hspace{-2mm}
	}
	\foreach \x in {A, ..., Z}{\expandafter\xdef\csname orcid\x\endcsname{\noexpand\href{https://orcid.org/\csname orcidauthor\x\endcsname}
			{\noexpand\orcidicon}}
}

\begin{document}
\title{Johann Bernoulli's analysis of elastic collisions\\
(a teaching sequence  to introduce the  dynamic law  thereby inspired)} 

\author{Roberta  T. Campos,$^1$\footnote{trigueiro.beta@gmail.com}
Mariana  F.  B. Francisquini,$^2$\footnote{mariana.francisquini@ifrj.edu.br
} \\
P. M. Cardozo Dias$^3$\footnote{penha@if.ufrj.br}\orcidA{}
\\
$^1$Col\'egio da Pol\'icia Militar do
Estado do 
\\
Rio de Janeiro, Niter\'oi, Brazil
\\
$^2$Instituto Federal de Educa\c{c}\~ao, Ci\^encia e Tecnologia 
\\ do Rio de Janeiro,  Niter\'oi, Brazil
\\
$^3$Instituto de F\'isica,  Universidade Federal 
do Rio de Janeiro, 
\\
Rio de Janeiro, Brazil}

\date{}
\maketitle
\begin{abstract}
In order to explain an elastic collision, Johann Bernoulli considered two  bodies  connected by a spring.  Motion is defined as a succession of states of rest. Then, considering the spring to  be a lever  with a body at each extremity,  the laws of equilibrium imply  that `motion' is described by the time variation  of the (common) quantity of motion of the bodies; the dynamic law is thus deduced. 
This inspires a teaching sequence to introduce the  dynamic law  (in one dimension) in introductory physics course; we call it  ``bernoullian 
sequence''.
\end{abstract}

\noindent{\it Keywords}:
Newton's  second law,  teaching sequence,  physics education
%%%%%%%%%%%%%%%%%%%%%%%%%%%%%%%%%%%%%%%%%%%%%%%%%%%%%%%%%%%%%%%%%%%%%%%%
\section{Introduction}
\label{sec:int}
A survey of  books  used in introductory physics courses  in the city of Rio de Janeiro, Brazil,  
discloses that  in those books the dynamic law of motion is introduced acccording to either of the two teaching sequences identified by Arnold Arons \cite{Roberta}.
These are a sequence that Arons associates with  Ernst Mach, and another sequence that he associates with Isaac Newton \cite{Arons}. 
 In the machian sequence \cite[p~58]{Arons}:
\begin{quote}
{\footnotesize 
[\dots] inertial mass is defined first. This is done by invoking
the reaction car experiment,\ignorespaces
\footnote{``The ``reaction car'' experiment is that
in which two carts on a table, or gliders on an air track, have a compressed spring
between them and fly apart when the spring is released" \cite[p~42]{Arons}}
%%%ENDFOOT
accepting as a law of nature the empirical observation
that the ratio of the accelerations (and hence of the velocity changes)
of the two bodies is a fixed property of the bodies, and defining the ratio of
the masses as the inverse ratio of the accelerations. The net force acting on
one body is then defined as the $ma$ product for that body.}
\end{quote}
In the newtonian sequence \cite[p~42]{Arons}:
\begin{quote}
{\footnotesize 
[\dots] that is, starting with force and acceleration
rather than with Mach's reaction car experiment. This, however, is still not the entire content of the second law. There remain the questions of superposition
of forces and masses, and again one must appeal to experiment for verification of conjectures, however plausible the latter might be.}
\end{quote}
In those approaches akin to the newtonian sequence (whether the book is aimed to high school students or to more advanced introductory courses),  the quantity of motion is artificially introduced, and justification of the law consists rather in descriptions of how it works; when history is invoked as support to the law, it is not faithful to documentary  evidences. Only one of the surveyed books follows a machian approach; the quantity of motion is introduced  by actual experiments, but  the experiments  involve sophisticated graphs, which makes the approach of little help in mass education, in spite of the excellent   textbook material.  

What physicists  call (not  correctly on the point of view of history) dynamic equation of  motion    is the Axiom II,  stated in 1686 by Isaac Newton  in   {\it Philosophi{\ae} Naturalis Principia Mathematica}  \cite{Newton}. However, the  axiom was not  introduced as a law of motion;
Isaac  B. Cohen argues that,  as it stands in the {\it Principia}, Axiom II  is better understood as a definition rather than a dynamic law \cite[p.~158]{Cohen1970}. Furthermore, as  late as 1744, Jean Le Rond D'Alembert, in his {\it Trait\'e de Dynamique} \cite{Dalembert},  presents arguments against  the idea of an equation of motion, and against the concept of `force'; accordingly, an equation of the form ``cause $=$ effect" is an axiom  ``vague and obscure" \cite[p.~xi]{Dalembert}, and  the force in the expression $F=ma$ is not a separate entity, but a name for $ma$ \cite[p.~25]{Dalembert}.
 Hindsight not withstanding, D'Alembert's criticisms raise questions that pertain to the foundations of the categories of mechanics: why should the dynamic law be expressed by a(n) (differential) equation that connects a cause to its effect? Why a second order derivative of time? After all, taking for granted the law of inertia, it at most states that whatever changes velocities is a function of the acceleration, but does not state what function it is, nor  what  ``whatever" means.
%; why not 
%$|\vec{F}|\propto \sqrt[3]{a^2}$, 
%or $\vec{F}\propto\frac{d\vec{a}}{dt}$? 

Deductions of the dynamic law in the eighteenth century give answers to D'Alembert's criticisms, even if they  do not directly and textually   refer to those criticisms.
In an approach,  a dynamic principle  is given by    Galileo Galilei's theorem on the fall ($v^2\propto 2h$, or $h\propto t^2$);  general motions result  from the application of the theorem to infinitesimal (virtual) motions. 
So  did: (1)
Christiaan Huygens \cite{Eu2013}, in {\it Horologium Oscillatorium} \cite{HuyHorOsc} and in {\it De Vi Centrifuga} \cite{HuyVi},  to find the centrifugal 
force; (2)
Newton \cite{Eu2023},  to describe planetary orbits, in the {\it Principia} \cite{Newton};
 (3)
Leonhard Euler  \cite{Eu2017},  to demonstrate the differential  equation of motion of a point mass, in   {\it  Recherches} \cite{EulerRecherches} and in 
{\it D\'ecouverte} \cite{EulerDecouverte}; (4)
 Euler again \cite{Eu1999},   to introduce the methods of the nowadays called analytical mechanics, in {\it R\'eflexions}  \cite{EulerReflexions} and   in {\it Harmonie} \cite{EulerHarmonie}.
A different approach is found in Johann
 Bernoulli's {\it Discours sur les Loix de la Communication du Mouvement}  \cite{JohannB};    motion is a succession of states of rest;
dynamics is   introduced by 
applying the laws of statics to infinitesimal (virtual) motions.

Johann Bernoulli's discussion of  elastic collisions
produces a deduction of the one-dimensional
 dynamic equation  (section~\ref{sec:tenets}). This approach to dynamics inspires a teaching sequence to introduce the second law in elementary physics courses (section~\ref{sec:material}); we call it ``bernoullian sequence''.  It  has pros and cons vis-\`a-vis the newtonian and machian sequences, and shares with them some  difficulties  (section~\ref{sec:prosandcons}). 
However, in the bernoullian sequence, the dynamic law results from a deduction which  gives a meaning to its constitutive terms and its functional form (section~\ref{sec:relevance}).  
We also comment  on the ancillary role that  case studies may play in  physics education (section~\ref{sec:ancillaryuse}). 
Not to overload the text, quotations of  primary sources are presented in  appendix~\ref{ap:quotations}, in the order they appear in the text

\section{Bernoulli's treatment of dynamics}
\label{sec:tenets}
 %Bernoulli's treatment of  an accelerated motion  depends on a concept of rest or equilibrium, and on a concept of motion.
%\subsection{Description of rest}
%\label{sec:rest}
In Bernoulli's account, a body is kept at rest by the continual action of a force (he calls it ``dead force" after Gottfried Wilhelm Leibniz) that opposes an ``external obstacle" \cite[ch.~V, \S2, p.~32]{JohannB}.
%\ignorespaces
%\footnote{``The dead force consists in a simple effort: this effort can subsist, whenever  an obstacle external to the body prevents [the dead force] to produce at any moment a local motion of the body on which the effort displays itself". \cite[p~35]{JohannB} [Free translation].}
%%%
Rest  consists in a continual,  virtual process of ``creation'' of motion followed by  its ``destruction'', before motion  sets in.
This is illustrated  
by a body resting on a table: at each instant, the body tends to fall, but the table is an opposing obstacle, so that the weight produces a dead force upwards that lasts during only an instant; 
%\ignorespaces
%\footnote{``A heavy body resting  on a horizontal %table makes a continual effort to move down. It would really move down, if the table were not an opposing obstacle that retains [the body]. In this way, the weight produces a dead  force whose effect is only momentary".  \cite[p~35]{JohannB}[Free translation].}
%%%%%%%%%%%%%
accordingly,  equilibrium is described as a continual,  virtual process of ``creation'' of motion followed by  its ``destruction''  
\cite[ch.~V,  \S2, p.~32]{JohannB}.\ignorespaces
%\ignorespaces
%\footnote{``At each instant, the weight imprints an infinitesimal  degree of speed on bodies on  which it acts; this degree is immediately absorbed by the resistance of the obstacle. These small degrees of speed are destroyed  at birth and are reborn at death. When the weight is constrained by an insurmountable  obstacle, its effort consists in that constant reciprocation of production and destruction, to which we have given the name of dead force. As to the obstacle, when it resists to the effort of the weight, it receives from this pressure a force always equal and reciprocal to the force due to the action of the weight. The dead force [the force that keeps equilibrium by cancelling external forces] has this particularity: it does not produce an effect that lasts longer than it; if the force ceases, everything ceases together with it, and its effect never  survives its action; if a heavy body on a table suddenly looses its weight, then at the same instant the table ceases to be pressed".  \cite[p~35-36]{JohannB} [Free translation].}
%%%%%
\footnote{The idea of ``virtual" quantities, and the law of virtual velocities had been used since  Antiquity to describe static equilibrium \cite[in many places in the book]{Clagett}.}
%\ignorespaces
%\footnote{``Hypothyesis: Two agents are in equilibrium, or have equal moments, when their absolute forces are in a reciprocal ratio to their virtual velocities, whether the  forces that act on each other are in motion or at rest". \cite[p.~35]{JohannB}[Free translation].}
%\subsection{Description of motion}
%\label{sec:motion}

`Motion' is understood as succession of rest or equilibrium states. An example is given by an elastic collision between two bodies; elasticity is realized  by a spring  connecting the bodies (figure~\ref{fig:fig5} below).  According to Bernoulli, 
``the two opposed efforts of the spring being 
equal", it forms a lever in equilibrium,\ignorespaces 
\footnote{That the internal forces in a system form a lever in equilibrium had already been  stated  by Jakob Bernoulli, Johann's elder brother \cite{JakobB}.}
%%%%
with the masses at the extremities, 
respectively.\ignorespaces
%%%%
\footnote{In the case of a lever at rest (figure~\ref{fig:fig1} below),  as the weight on a side of a lever ``creates'' a tendency to move the side down,
the weight on the other side opposes this tendency.}
%%%%%%%%%%%%%%%%%%%%%%
The virtual  motion  is described  by  the ``principle of virtual velocities"  \cite[ch.~III, \S3, p.~20]{JohannB}.
%from the hypothesis taken from mechanics [the law of virtual velocities] that the force of inertia of $A$ is to  the force of inertia of $B$ --- or that the mass of $A$ is to the mass of $B$ --- in a ratio reciprocal to the ratio of the virtual velocity of the body $B$ to the virtual velocity of the body $A$" \cite[ch.~III, \S3, p.~20]{JohannB}; 
%\ignorespaces
%\footnote{ ``Suppose two bodies at rest, $A$ et $B$, and between them a strained spring, $C$. As the spring stretches, it makes an equal effort on the bodies, so that the bodies $A$ and $B$ move away from each other. It is clear that due to its inertia, each body opposes a resistance to the motion of the spring, which is  proportional to its mass. [\dots]. The two opposed efforts of the spring being equal, it is necessary from the hypothesis taken from mechanics that the force of inertia of $A$ is to  the force of inertia of $B$ --- or that the mass of $A$ is to the mass of $B$ --- in a ratio reciprocal to the ratio of the virtual velocity of the body $B$ to the virtual velocity of the body $A$.  \cite[p~24]{JohannB} [Free translation].}

The dead force  is also  the force that causes a body at rest to move, or to change its speed, if it is already moving \cite[ch.~III, Definition II, p.19]{JohannB}; in fact, if the dead force keeps a body at rest, then by removing it the body starts to move under the action 
of  an unbalanced ``obstacle" equal to the dead force. Then:
\begin{equation}
\mbox{measure of dead force}= 
\frac{\delta}{\delta t}\left(mv\right).
\label{eq:deadforce}
\end{equation}
This discussion can be summarized in the form of the syllogism in table~\ref{tab:silogismo};  the conclusion  is the  dynamic equation of motion.
Concepts such as mass and force seem to refer to the leibnizian philosophic context.\ignorespaces
%%%%%%%%%%%%%%%%%%%%%%%%%%%%%% 
\footnote{Leibniz introduces  four kinds of force:
 primitive active, primitive passive, derivative active (subdivided in dead and living), derivative passive. Leaving aside  philosophic rigor and hindsight,
primitive passive force is better understood as mass, and derivative active forces are, respectively, the forces in physics (dead) and $mv^2$ (living).} 
\begin{table}[!h]
\caption{Syllogism that gives meaning to the dynamic law}
\label{tab:silogismo}
\begin{tabular}{l|c}
\\
the concept of  equilibrium implies that
&
{\bf cause of a motion}
\\
the cause of a motion is measured by
&
%%%%
{\bf that lasts for} {\boldmath $\delta t$}
\\
a force that opposes motion (dead force)
&
{\boldmath$=$}
{\bf dead force} {\boldmath{$\times \delta t$}}
\\
\hline
the law of equilibrium implies that the
&
{\bf effect of motion}
\\
effect of motion is the creation of {\boldmath$\delta\left(mv\right)$}
&
{\boldmath$=\delta\left(mv\right)$}
\\
\hline
metaphysical principle:
&
cause
$=$
effect
\\
\hline
conclusion:
&
{\bf dead force} {\boldmath$\times \delta t$}
\\
&
{\boldmath$
=\delta\left(mv\right)$}
\\
\end{tabular}
\end{table}

%%%%%%%%%%%%%%%%%%%%%%%%%%%%

\section{The ``bernoullian sequence''}
\label{sec:material}
In the bernoullian sequence, the ideas to be conveyed to the students are: 
\cite{Roberta}:
\begin{enumerate}
\item 
The laws of rest. These are:
\begin{itemize}
\item
The law of the lever.
\item 
The law of virtual velocities. 
\end{itemize}
\item
The definition of motion as a succession of states of rest. 
 \end{enumerate}
 
\begin{figure}[!!h]
\centering
\includegraphics[height=5cm]{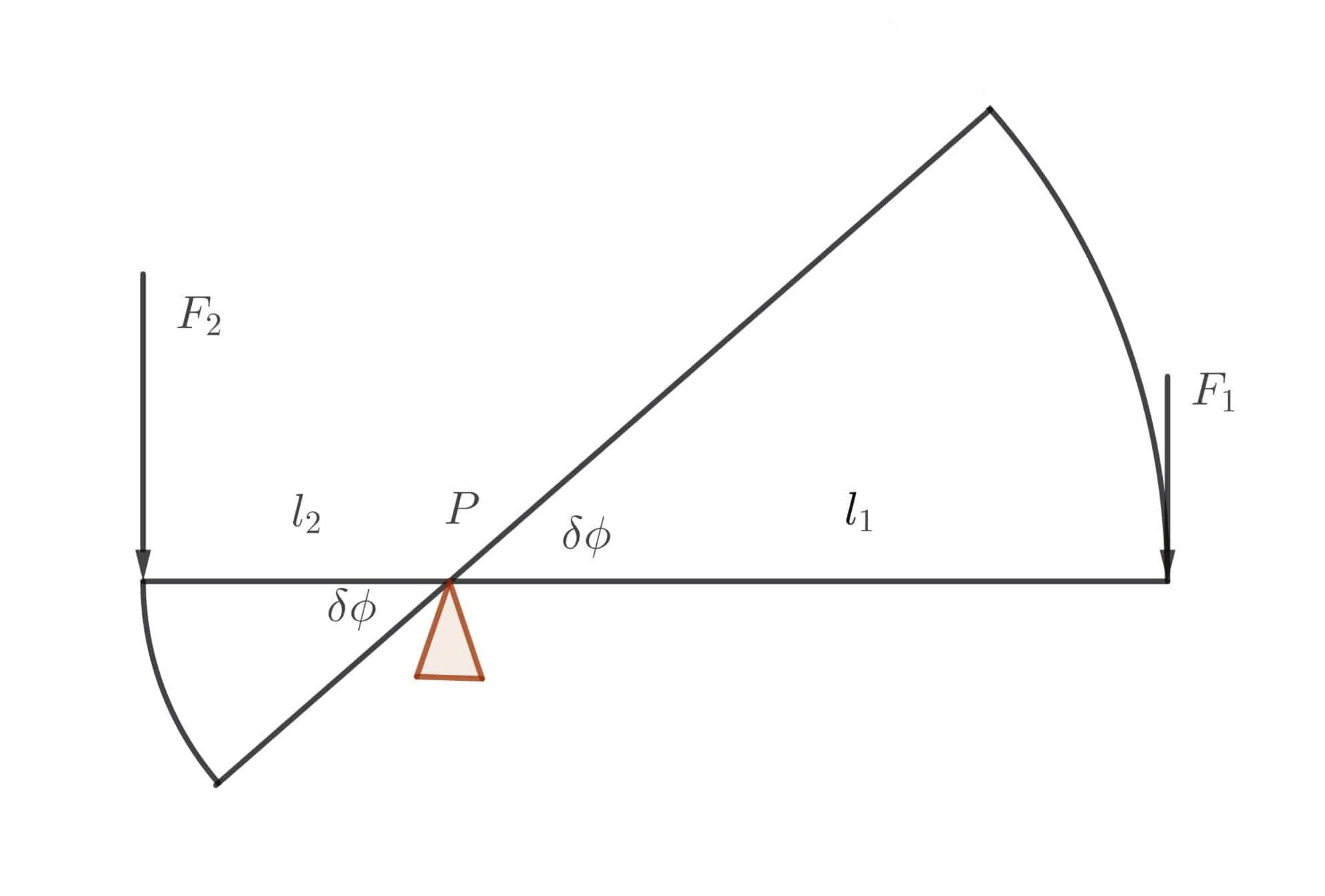}
\caption{The lever. When equilibrium is broken, the lever rotates $\delta\phi$ around the fulcrum.}
\label{fig:fig1}
\end{figure}

In figure~\ref{fig:fig1}, 
  $P$ is the fulcrum; $F_1=m_1g$, $F_2=m_2g$  are weights at the extremities of the arms; $l_1$, $l_2$ are the lengths of the arms. 

When equilibrium is broken, the arms  rotate in an infinitesimal time  an angle $\delta\phi$ around the fulcrum, while the extremities describe arcs of lengths $\delta s_1$ and $\delta s_2$; the (virtual) speeds 
of the  motions on the arcs are  are $\nu_1\equiv\delta v_1$, $\nu_2\equiv\delta v_2$.  Equilibrium of the lever  is equivalently described  by two principles:\ignorespaces
\footnote{The two principles are equivalent:
$$\frac{m_1}{m_2}=\frac{l_2}{l_1}\equiv
\frac{l_2\times\delta\phi}{l_1\times\delta\phi}=
\frac{\delta s_2}{\delta s_1}=\frac{\nu_2\delta t}{\nu_1\delta t}=\frac{\nu_2}{\nu_1}
$$
}

\begin{equation}
\mbox{principle of the  lever:} \quad \frac{m_1}{m_2}=\frac{l_2}{l_1};
\label{eq:eqlever}
\end{equation}
\begin{equation}
\mbox{principle of virtual velocities:}\quad
\frac{m_1}{m_2}=\frac{\nu_2}{\nu_1}.
\label{eq:virtualvel}
\end{equation}

  \subsection{The law of the lever}
 Equation~\ref{eq:eqlever} can be introduced to the students in many ways.   In \cite{Roberta} the law is introduced by simulations using the platform PhET.  
In the platform, objects of the same kind (for instance,  either garbage bin or fire extinguisher) are differentiated by the values of their masses.

\subsection{The  principle of virtual velocities
%: the description of equilibrium as a continual destruction  of ``would be'' motions
}
In equation~\ref{eq:virtualvel},  
 the $\left(\nu\right)$s are speeds that would be acquired, if the lever started to swing. 
 The principle is introduced in two steps. In the first, the concept of ``virtual''  motion is elaborated. In the second, the law itself is proved, which involves an elementary algebraic  calculation. 
\subsubsection*{Step 1: the idea of ``virtual'' motion}
Although the idea of ``virtual'' quantities that exist in ``would be'' infinitesimal  motions ($\nu\equiv\Delta v$)  
is difficult to a student in an introductory level, 
 it is possible to give a pictorial description of ``virtual'' motion, if one leaves aside mathematical rigour.   One meets the very same difficulty, 
  when teaching instantaneous velocity or acceleration.
  First, it is proposed an exercise; from it,  the virtual $\Delta v$ is defined.

\subsubsection*{Exercise}
\begin{figure}[!h]
\centering
\includegraphics[width=7cm]{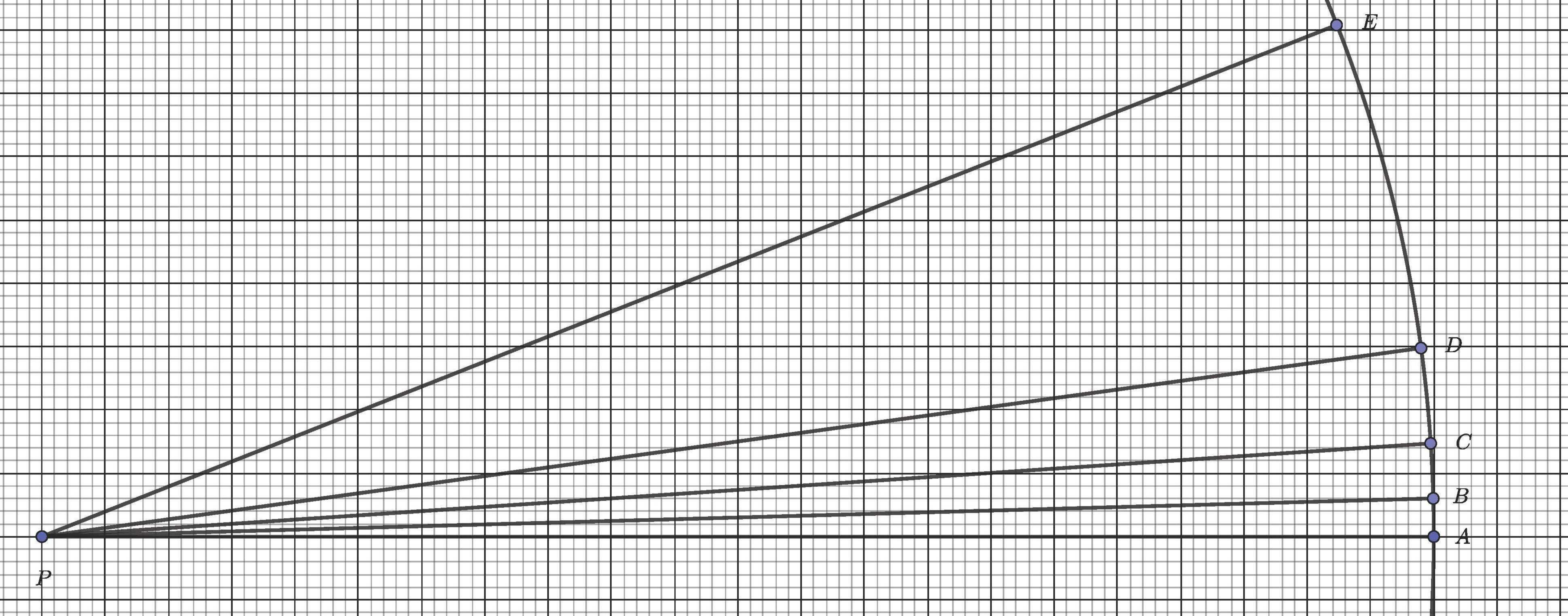}
\caption{Positions of the lever arm. The figure shows the position of an arm of the lever for different angles 
of rotation around the fulcrum ($P$). Arc $AB$ conflates with a straight line}
\label{fig:fig2}
\end{figure}

\begin{figure}[!h]
\centering
\includegraphics[height=4cm]{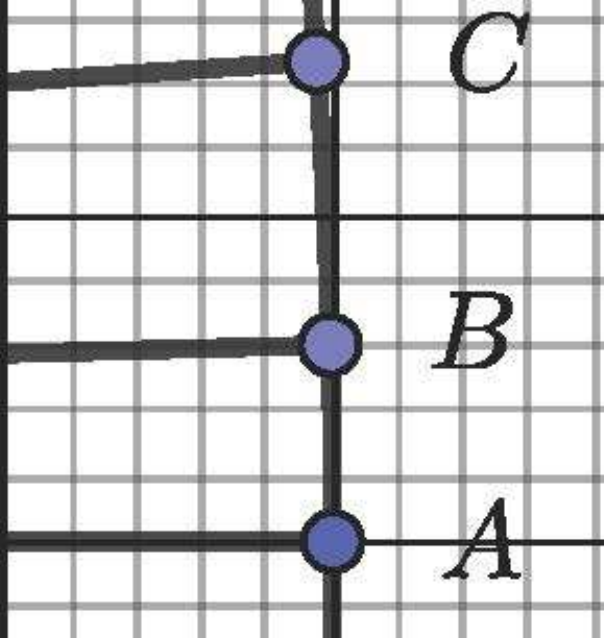}
\caption{The arc for ``small'' angles. Segment $AB$ coincides with the line of the mash.}
\label{fig:fig3}
\end{figure}

Students are requested to draw  on a graph paper the position of the lever arm   for angles that become smaller and smaller (figure~\ref{fig:fig2}). They have to observe that as the angles becomes smaller, the arc becomes a straight line: taking a photo with a smart phone, so that the figure can be enlarged (figure~\ref{fig:fig3}), it is observed that the arc $AB$ coincides with a straight line. Angles such that the arc becomes straight introduce the idea of a  ``small angle''.

\subsubsection*{Definition of the  virtual speed}
If the lever started to rotate, a body at its 
the extremity  would  acquire a speed $\Delta v$ after a time $\tau$ during which the arc coincides with the straight line. The straight line segment supports an uniform speed (law of inertia), so that $\Delta v$ is the mean speed acquired in $\tau$ starting from rest ($\Delta v=v-0$). The virtual speed is defined by the mean speed on the straightened arc. 

\subsubsection*{Step 2: The law itself}
\begin{figure}[!h]
\centering
\includegraphics[width=8cm]{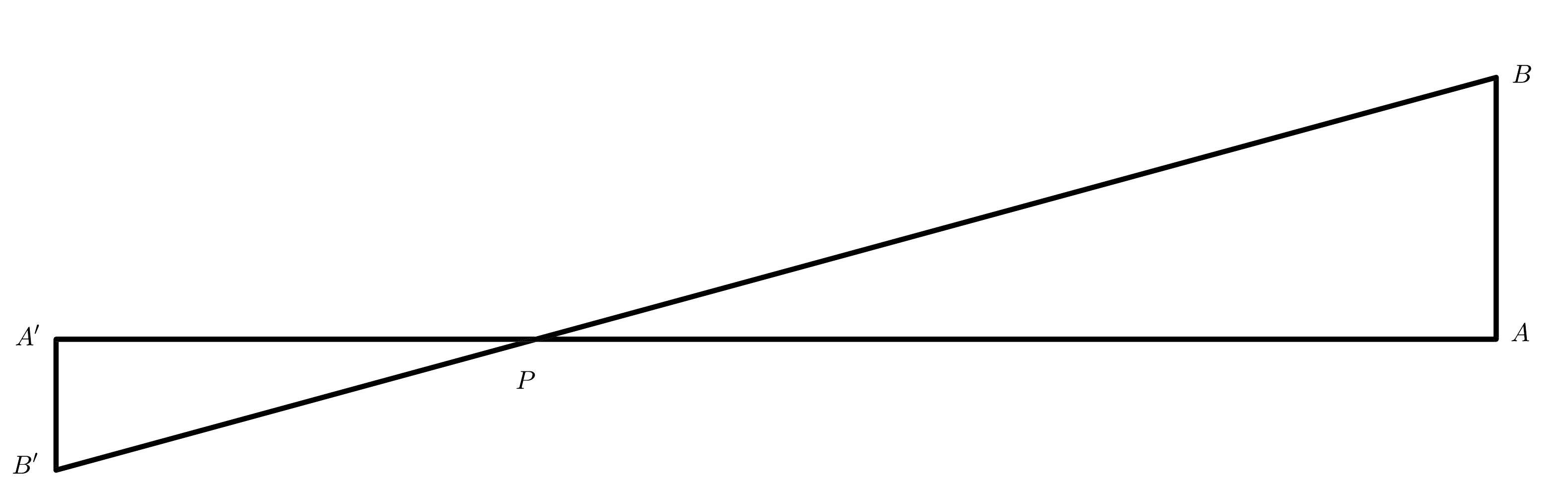}
\caption{The two similar right triangles formed in a rotation of $AA^{\prime}$ by a small angle.}
\label{fig:fig4}
\end{figure}

For small angles, the motion of the lever forms two similar right triangles (figure~\ref{fig:fig4}). 

From the similarity:
\begin{equation}
\frac{PB}{PB^{\prime}}=\frac{PA}{PA^{\prime}}=\frac{AB}{AB^{\prime}}
\label{eq:proporcoes}
\end{equation}
Now, $PA=l_1$, $PA^{\prime}=l_2$, $AB=\Delta v_1\tau$, $A^{\prime}B^{\prime}=\Delta v_2\tau$, so that equation~\ref{eq:proporcoes} can be written:
$$
\frac{l_1}{l_2}=\frac{\Delta v_1\tau}{\Delta v_2\tau}\quad{\mbox{or}}\quad \frac{l_1}{l_2}=\frac{\Delta v_1}{\Delta v_2}$$

%%%

Putting together equation~\ref{eq:eqlever} and equation~\ref{eq:virtualvel}, the law of equilibrium is:
\begin{equation}
\frac{m_1}{m_2}=\frac{\Delta v_2}{\Delta v_1}
\end{equation}
or
\begin{equation}
m_1\Delta v_1=m_2\Delta v_2
\label{eq:momentumrest}
\end{equation}
Equation~\ref{eq:momentumrest} means that 
 the two bodies remain at rest, if 
they acquire  in an infinitesimal time an equal value of the quantity $P=mv$ --- the ``quantity of motion"  --- but in opposite directions. 

\subsection{Motion as a succession of states of rest}
\begin{figure}[!h]
\centering
\includegraphics[width=6cm]{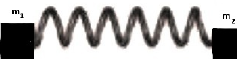}
\caption{The motion of the oscillator represents an elastic collision between two masses}
\label{fig:fig5}
\end{figure}

The spring in figure~\ref{fig:fig5}  is  stretched (or compressed), and then released; 
 the masses move on a horizontal line, successively compressing and stretching  the 
spring.\ignorespaces
\footnote{
The solution of this problem is: 
the masses oscillate out of phase (i.e., either both compress or both stretch the string) with equal frequency but different amplitudes, i.e., 
$x_1=\Lambda\cos\left(\omega t+\Phi\right)$ and $x_2=\frac{m_1}{m_2}\Lambda\cos\left(\omega t+\Phi\right)$; $\Lambda$ and $\Phi$ are constants of integration, 
$\omega=\sqrt{\frac{k}{\mu}}$ is the  frequency of oscillation,  $\mu=\frac{m_1m_2}{m_1+m_2}$ is the reduced mass,  $k$ is the spring constant.} 

The motion  is shown  in an animation by physics-animation.com:
\begin{center}
\url{https://www.youtube.com/watch?v=9XmnB_y_gi4}
\end{center} 
A video of the actual experiment   is found in:\ignorespaces
\footnote{Between the instants  $34$ seconds and $50$ seconds, the experiment  is filmed from afar; between the instants  $1$ minute and  $32$ seconds,  and   $1$ minute and $50$ seconds, it is filmed closely.}
\begin{center}
\url{https://www.youtube.com/watch?v=CjJVBvDNxcEt=0s}
\end{center}

By placing  obstacles to the  motion of the masses,
one can annihilate  the action of the spring on each mass, and  
the system mass-spring-mass is brought into 
rest.\ignorespaces
\footnote{ 
In digital life, the animation can be frozen at any position by  a click on the pause icon. In real life,  
 there must be a real obstacle, v.g., one holds the masses.}
%%%%%%%%%%%%
 The
 law of virtual velocity  (equation~\ref{eq:momentumrest}) 
applied to  each instant separately is shown in  
table~\ref{tab:instantaneousmomentum}.

\begin{table}[h]
\centering
\caption{Motion of the oscillator as a succession of states of rest}
\label{tab:instantaneousmomentum}
\begin{tabular}{l|c}
\hline
Instant 
&
Mean quantity of motion that a mass 
would acquire,
\\
&
 if it started to move at the instant
\\
\hline
\\
$t_0$
&
$\left[\Delta\left(m_1v_1\right)\right]_{t_0}= \left[\Delta\left(m_2v_2\right)\right]_{t_0}$
\\
$t_1$
&
$\left[\Delta\left(m_1v_1\right)\right]_{t_1}= \left[\Delta\left(m_2v_2\right)\right]_{t_1}$
\\
$t_2$
&
$\left[\Delta\left(m_1v_1\right)\right]_{t_2}= \left[\Delta\left(m_2v_2\right)\right]_{t_2}$
\\
\dots
&
\dots
\\
$t_n$
&
$\left[\Delta\left(m_1v_1\right)\right]_{t_n}= \left[\Delta\left(m_1v_1\right)\right]_{t_n}$
\\
\dots
&
\dots
\\
$T$
&
$\left[\Delta\left(m_1v_1\right)\right]_{t_T}= \left[\Delta\left(m_2v_2\right)\right]_{t_T}$
\end{tabular}
\end{table}

%%%%%%%%%%%%%%%%%%%%%%%%%%%%%%%%%%%%%%%%%%%%%%%%%%%

\subsubsection*{Consequence}
\begin{enumerate}
\item
From the law of virtual velocities, in a frozen instant the masses acquire (or loose) an equal value of the quantity of motion ($\Delta P=\Delta\left(mv\right)$). Then (table~\ref{tab:instantaneousmomentum}):
\begin{itemize}
\item
The accelerated motion of each  mass  is characterized by the time variation of its quantity of motion.
\end{itemize}
\item
This implies:
\begin{itemize}
\item
To   keep any one of the masses at rest, the time variation of its quantity of motion must be    annihilated  by whatever is preventing the motion of the mass (call it $f$): 
$\frac{\Delta\left(mv\right)}{dt}-
\left(\mbox{measure of $f$}\right)=0$
\end{itemize}
\item
Whatever   prevents the  motion of   the mass ($f$) must annihilate  whatever  attempts  to create  a change of its  quantity of motion, the push or pull of the string on the masses   ($F$): 
$\left(\mbox{measure of $f$}\right)-\left(\mbox{measure of $F$}\right)=
0$.\ignorespaces
\footnote{This is  Newton's third law. It can be separately stated, if students already know it, or the conception is worked, when the sequence is applied in classroom.} 
\item
Conclusion: the push or pull of the string on the masses is
\begin{equation}
F=\frac{\Delta\left(mv\right)}{\Delta t}.
\label{eq:leiformal}
\end{equation}
\end{enumerate}

\section{Pros and cons 
\\ (vis-\`a-vis the newtonian and machian sequences)}
\label{sec:prosandcons}
Arons criticizes the use made of both sequences in textbooks. In the case of the newtonian sequence   \cite[p~65]{Arons}:
\begin{quote}
{\footnotesize
The majority of widely used textbooks seem to adopt what I have called the ``Newtonian sequence,'' but most of these start with ``force'' as though it were a primitive, already fully understood both qualitatively and numerically, and not requiring explicit operational definition. They then go on to ``mass'' as simply the proportionality constant between force and acceleration.}
\end{quote}
In the case of the machian sequence \cite[p~65]{Arons}:
\begin{quote}
{\footnotesize 
In most texts adopting the Mach sequence, the presentation is made so cryptically and so abstractly as to be quite meaningless to the majority of students, even though the conceptual development is sound and not circular. [\dots].
The verbal text, the qualifications and interpretations that accompany the second law, are entirely omitted. The more ``elementary'' the textbook, the more cryptic and less intelligible is likely to be the
presentation.}
\end{quote}

The concepts of `mass' and `force'  can be considered to be primitive concepts in Bernoulli's calculations (they seem to be borrowed from a leibnizian philosophic context);   Arons's criticism to the newtonian sequence may therefore apply. Differently  from the newtonian  sequence,  the  bernoullian  sequence  does not intend to exhaust all there is in the law of motion;  to start with, the sequence only produces  the law in one dimension.

The bernoullian sequence starts with an elastic collision; as does the machian sequence. 
 Differently  from the machian sequence,  the  bernoullian  sequence does not   ``[accept] as a law of nature the empirical observation that the ratio of the accelerations [\dots] of the two bodies is a fixed property of the bodies", nor introduces the ratio of masses as a definition; in the bernoullian sequence the equality of the variation of the quantities  of motion of the two bodies   is justified on more general principles. Furthermore, the machian sequence relies on actual  experiments with the ``reaction car"
  to find the ratio of accelerations, which  involves measurements of  quantities that are a-priori  given,  graphs, etc., and   introduces more difficulties and  abilities.

Finally, the bernoullian sequence shares with any other introductory  approach to the dynamic law the  difficulty of making sense of   quantities that are obtained by taking the derivative   of other quantities;  
this is the case of instantaneous velocity   and of instantaneous acceleration \cite{Trowbridge}.  By the same token,  
it is difficult to give mathematical precision to  equation~\ref{eq:leiformal}.

\section{All in all:  relevance of the bernoullian sequence}
\label{sec:relevance}
In the bernoullian sequence,   the  equation of 
motion is actually deduced from the  characterization of motion as a succession of states of rest (table~\ref{tab:silogismo}): 
\begin{itemize}
\item
The quantity of motion  naturally arises: its equality in both bodies is a necessary condition for rest. 
A sequitur is that the functional form of the dynamic law must necessarily involve the time variation of the quantity of motion.
\item
This is discussed with the help of  a simulation  and  an animation.  These can be taken as 
 thought experiments. Actual experiments may not be   free from intrusive elements of reality that
 are not part of the idea to be conveyed, and may divert the mind of the students to experimental contingencies.  Thought experiments, on the contrary \cite[p~167]{MatthewsLivro}:
\begin{quote}
{\footnotesize  
These types of  `thinking physics' problem allow teachers and students to determine what they mean by fundamental concepts such as gravity, force, pressure and so on, and to think about the correct conditions for the applicability of concepts.}
\end{quote}
\end{itemize}

In the simulations with the lever,  masses are given, not their weights. This conflates two different concepts, weight and mass. However there is no harm in doing so:  the acceleration of gravity is canceled in equation~\ref{eq:eqlever} and in equation~\ref{eq:virtualvel}, making it possible to obtain equation~\ref{eq:momentumrest} without discussing the nature of masses and forces;  anyhow this conflation is quite  unavoidable without the dynamic law, and both  concepts can be reviewed.  
There certainly is a great gain: the quantity of motion  naturally appears in equation~\ref{eq:momentumrest}.

The bernoullian sequence presents  a 
derivation of the dynamic law; 
it  can be understood as a quite qualitative  argument to the effect that   an accelerated motion  is described by the time variation of the quantity of motion, and that this variation results from an unbalanced  push or pull of the spring on the mass.\ignorespaces
\footnote{When applying the sequence in  classroom, it may be necessary to discuss the conception of ``canceling". In \cite{Minstrell} the  concept of equilibrium is investigated;  it is asked to  high school students: ``what keeps a book at rest on a table?" Students presented many ingenious answers, but missed the force of the table on the book, and in some answers also the idea of  ``cancellation" of forces in a situation of rest.}

\section{An ancillary role of the history of physics in physics education}
\label{sec:ancillaryuse}
The use of the history of physics in physics education is a long debated subject \cite{Matthewspaper}. Against the point of view that selectivity of episodes  in history can lead to distorted simplification,  Helge Kragh claims that selective history can be done, provided  
 ``[\dots]  it does not serve ideological purposes or violates knowledge of what actually happened" 
\cite[p.~360]{Kragh}.

On this pragmatic point of view, cases studies
are  ``archaeological sites", that can be  excavated
in the search of the meaning and the justification of the categories of physics. Early formulations of  laws or of any concept and idea  disclose   the  ``whys and wherefores" and the ``hows", pointing to those meanings and justifications  
on which laws and concepts were first constructed; perhaps, because early formulations are free from conceptions  that may  appear later, when laws and concepts are inserted in  wider contexts. 
In this sense, cases studies are more relevant in learning the foundations of physics   than in teaching physics; 
yet, case studies may be explored in physics 
education. 
The best way to explain this ancillary role of case studies   is by examples:
\begin{itemize}
\item
	Galileo uses the medieval ``double distance rule'' together with the laws of uniform motion to deduce that in a fall $v^2\propto 2h$ (\cite{Eu2019},  2019). 
Galileo's reasoning led to  a proposal to teach the concept of instantaneous speed; this proposal was   explored  in two different experiments, respectively in \cite{Carlos} (2022)
and in  \cite{Glauce} (2022).\ignorespaces
\footnote{In \cite{Glauce},  it is presented a  teaching sequence, in which the experiment is rather a tool to convey ideas, as in the bernoullian sequence.}
\item
Newton's approach to orbital motion starts from the
drawing of the orbit by the motion of a point: the orbit is the composition of two virtual motions, an uniformly accelerated motion to the center of force, and an uniform motion along the tangent. 
Newton's drawing method leads to a proposal to obtain the law of force in an  elliptic orbit that can  be used in  physics education \cite{Rojans} (2022).
The drawing was made using the GeoGebra, but  any drawing method could be used, or one could draw  the orbit by hand, as did Newton in a letter to Hooke, on December 13, 1679  (\cite{Eu2023}, 2023).
\item
In this paper,  a new item is added to the list.
\end{itemize}

\section{Round up}
\label{sec:roundup}
A case study in the history of physics 
discloses a demonstration of the dynamic law, the so-called newton second  law. 
The demonstration inspires  an approach to the  law that can be taken into classroom, in introductory courses. 

The bernoullian sequence does  not intend to be self-sufficient. Some of the involved conceptions may have to be  worked separately; nor does it contain everything involved in the law. Even so, students in introductory courses can benefit from a quite qualitative approach to the functional form of the law, when first presented to it.
Perhaps, one should take into account 
Arons's advise  \cite[p~58]{Arons}: start from ``unsophisticated'' presentations, and, then, ``spiralling back to more rigorous definitions  as [the students's] grasp of the overall structure grows in later contexts''.

\appendix
\section{Quotations}
\label{ap:quotations}
\phantom{.}

{\bf \cite[p.~158]{Cohen1970}.}
 I would submit that Newton's procedure in thus producing a form of the Second Law out of Def. VIII prior to the formal statement of Law II among the Axioms not only shows how he anticipated the Axioms or Laws of Motion in the prior Definitions. It equally demonstrates that any form of the Second Law is basically a kind of Definition.

{\bf \cite[ch.~V, \S2, p.~32]{JohannB}.}
The dead force consists of a simple effort: this effort can subsist, whenever
an obstacle external to the body prevents [the  effort] to produce at any moment a local motion of the body on which the effort displays itself.
[free translation by the authors].

{\bf \cite[ch.~V,  \S2, p.~32]{JohannB}.} 
 A heavy body resting  on a horizontal table makes a continual effort to move down. It would really move down, if the table were not an opposing obstacle that retains [the body]. In this way, the weight produces a dead  force whose effect is only instantaneous. 
At each instant, the weight imprints an infinitesimal  degree of speed on bodies on  which it acts; this degree is immediately absorbed by the resistance of the obstacle. These small degrees of speed are destroyed  at birth and are reborn at death.
When the weight is constrained by an insurmountable  obstacle, its effort consists in that constant reciprocation of production and destruction, to which we have given the name of dead force. As to the obstacle, when it resists to the effort of the weight, it receives from this pressure a force always equal and reciprocal to the force due to the action of the weight. The dead force has this particularity: it does not produce an effect that lasts longer than it; if the force ceases, everything ceases together with it, and its effect never  survives its action; if a heavy body on a table suddenly looses its weight, then at the same instant the table ceases to be pressed. [free translation by the authors].

{\bf \cite[ch.~III, \S2 Hypohesis I, p.~20]{JohannB}.} 
Hypothyesis. Two agents are in equilibrium, or have equal moments, when their absolute forces are in a reciprocal ratio to their virtual velocities, whether the  forces that act on each other are in motion or at rest. [free translation by the authors].

{\bf \cite[ch.~III, \S3, p.~20]{JohannB}.} 
Suppose two bodies at rest, $A$ et $B$, and between them a strained spring, $C$. As the spring stretches, it makes an equal effort on the bodies, so that the bodies $A$ and $B$ move away from each other. It is clear that due to its inertia, each body opposes a resistance to the motion of the spring, which is  proportional to its mass. [\dots].
The two opposed efforts of the spring being equal, it is necessary from the hypothesis taken from mechanics that the force of inertia of $A$ is to  the force of inertia of $B$ --- or that the mass of $A$ is to the mass of $B$ --- in a ratio reciprocal to the ratio of the virtual velocity of the body $B$ to the virtual velocity of the body $A$. [free translation by the authors].

{\bf \cite[ch.~III, Definition II, p.~19]{JohannB}.} 
Dead force is the force that a body at rest receives, when it is  forced and prompted to move, or to  move with more or less speed, if it is already in motion.  [free translation by the authors].
%%%%%

 %%%%%%%%%%%%%%%%%%%%%%%%%%%%%%%%%%%%%%%%%%%%%%%%%%%%%%%%%

%%%%%%%%%%%%%%%%%%%%%%%%%%%%%%%%%%%%%%%%%%%%%%%%%%
\end{document}